\documentclass[twocolumn,showpacs,preprintnumbers,amsmath,amssymb,prb]{revtex4}

\usepackage{graphicx}
\usepackage{dcolumn}
\usepackage{bm}
\usepackage{amssymb}

\begin{document}

\title{Magnetic properties of a two-dimensional electron gas strongly coupled to light}

\author{K. Dini$^{1}$}
\author{O. V. Kibis$^{2,3}$}\email{Oleg.Kibis@nstu.ru}
\author{I. A. Shelykh$^{1,3,4}$}
\affiliation{$^1$Science Institute, University of Iceland, Dunhagi
3, IS-107, Reykjavik, Iceland} \affiliation{${^2}$Department of
Applied and Theoretical Physics, Novosibirsk State Technical
University, Karl Marx Avenue 20, Novosibirsk 630073, Russia}
\affiliation{$^3$Division of Physics and Applied Physics, Nanyang
Technological University 637371, Singapore} \affiliation{$^4$ITMO
University, Saint Petersburg 197101, Russia}

\begin{abstract}
Considering the quantum dynamics of 2DEG exposed to both a
stationary magnetic field and an intense high-frequency
electromagnetic wave, we found that the wave decreases the
scattering-induced broadening of Landau levels. Therefore, various
magnetoelectronic properties of two-dimensional nanostructures
(density of electronic states at Landau levels, magnetotransport,
etc) are sensitive to the irradiation by light. Thus, the
elaborated theory paves a way to optical controlling magnetic
properties of 2DEG.
\end{abstract}

\pacs{73.21.Fg,73.23.-b}

\maketitle
\section{Introduction}
The study of a two-dimensional electron gas (2DEG) exposed to a
high-frequency electromagnetic field is one of the most excited
areas in the modern physics of nanostructures. The permanent
interest to this topic originates from rich fundamental and
applied capabilities of two-dimensional electron systems (see,
e.g., Refs.~[\onlinecite{FerryGoodnick,Davies,Nag}]).
Particularly, the magnetoelectronic properties of 2DEG subjected
to a microwave irradiation are actively studied during last years
\cite{Zudov_2001,Ye_2001,Durst_2003,Lei_2003,Vavilov_2003,Dmitriev_2005,
Hatke_2009,Konstantinov_2009,Dmitriev_2009,Dmitriev_2012,Konstantinov_2013,
Zudov_2014,Shi_2015}. However, the most attention on the subject
was paid before to the simplest case of weak electromagnetic field
which does not change electron states. Namely, the only effect of
the weak field is the field-induced electron transitions between
the unperturbed states. On the contrary, a strong electromagnetic
field can substantially mix electron states. As a result of this
mixing, the composite electron-field object ``electron dressed by
field'' (dressed electron) appears \cite{Scully,Cohen-Tannoudji}.
The light-induced renormalization of physical properties of
dressed electrons has been studied in various atomic systems
\cite{Autler_55,Scully,Cohen-Tannoudji} and condensed-matter
structures, including bulk semiconductors
\cite{Elesin_69,Vu_04,Vu_05}, quantum wells
\cite{Mysyrovich_86,Wagner_10,Teich_13,Kibis_14,Morina_15,Pervishko_15},
quantum rings \cite{Kibis_11,Kibis_13,Joibari_14}, graphene
\cite{Lopez_08,Oka_09,Kibis_10,Kitagawa_11,Kibis_11_1,Usaj_14,Glazov_14,Lopez_15,Kibis_16},
etc. In the present research, we develop the theory describing the
magnetic properties of dressed 2DEG and demonstrate that they can
be substantially modified by the dressing field.

The paper is organized as follows. In the second section, we solve
the Schr\"odinger equation for a 2DEG subjected to both a
stationary magnetic field and a high-frequency dressing field. In
the third section, the found solutions of the Schr\"odinger
problem are used to analyze various magnetoelectronic
characteristics of dressed 2DEG, including density of electron
states and magnetotransport. The last sections contain conclusions
and acknowledgments.

\section{Schr\"odinger problem for Landau levels in dressed 2DEG}

Let us consider a two-dimensional electron gas (2DEG) confined in
the $(x,y)$ plane, which is subjected to both a stationary
magnetic field, $\mathbf{B}=(0,0,B)$, directed along the $z$ axis
and a linearly polarized electromagnetic wave (dressing field)
propagating along the same axis $z$ (see Fig.~1). The Hamiltonian
of 2DEG reads as
\begin{figure}
\includegraphics[scale=0.2]{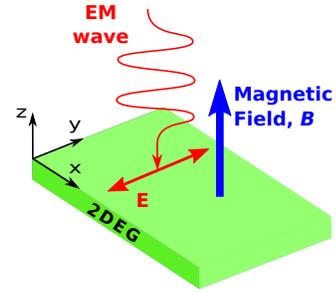}
\caption{\label{fig:figure1} (Color online) Sketch of the system
under consideration: Two-dimensional electron gas (2DEG) subjected
to both a linearly polarized electromagnetic wave (EM) with the
electric field amplitude, $E$, and a stationary magnetic field,
$\mathbf{B}$, directed perpendicularly to the 2DEG plane.}
\end{figure}
\begin{equation}\label{H2DEG}
\hat{\mathcal{H}}_e=\frac{1}{2m_e}\left[\hat{\mathbf{p}}-{e}\left(\mathbf{A}_0+
{\mathbf{A}}_t\right)\right]^2,
\end{equation}
where $m_e$ is the effective electron mass, $e$ is the electron
charge, ${\mathbf{A}}_0=(-By,0,0)$ is the stationary vector
potential of the magnetic field,
${\mathbf{A}}_t=(0,[E/\omega]\cos\omega t,0)$ is the
time-dependent vector potential of the electromagnetic wave, $E$
is the amplitude of electric field of the wave, $\omega$ is the
wave frequency, and $\hat{\mathbf{p}}=(\hat{{p}}_x,\hat{{p}}_y,0)$
is the operator of two-dimensional electron momentum, $p_{x,y}$.
Solutions of the nonstationary Schr\"odinger problem with the
Hamiltonian (\ref{H2DEG}) should be sought in the form
\begin{eqnarray}\label{psi}
\psi(\mathbf{r},t)&=&\frac{1}{\sqrt{L_x}}\exp\left[i\frac{p_xx}{\hbar}+i\frac{eE(y-y_0)}{\hbar\omega}\cos\omega
t\right]\nonumber\\
&\times&\phi(y-y_0,t),
\end{eqnarray}
where $L_{x,y}$ are dimensions of the 2DEG plane,
$\mathbf{r}=(x,y,0)$ is the radius-vector of electron in the 2DEG
plane, and $y_0=-p_x/eB$ is the center of cyclotron orbit along
the $y$ axis. Substituting the wave function (\ref{psi}) into the
Schr\"odinger equation with the Hamiltonian (\ref{H2DEG}),
$i\hbar\partial\psi/\partial t=\hat{\mathcal{H}}_e\psi$, we arrive
at the equation for the driven quantum oscillator,
$$
\left[\frac{m_e\omega_0^2y^2}{2}-eEy\sin\omega
t-\frac{\hbar^2}{2m_e} \frac{\partial^2}{\partial
y^2}-i\hbar\frac{\partial}{\partial t}\right]\phi(y,t)=0,
$$
which has the well-known exact solution (see, e.g.,
Refs.~[\onlinecite{Husimi,Haggi,Inoshita}]),
\begin{eqnarray}\label{phi}
\phi(y,t)&=&\chi_N(y-\zeta(t))\exp\left[-\frac{i\varepsilon_N
t}{\hbar}
+\frac{im_e\dot{\zeta}(t)[y-\zeta(t)]}{\hbar}\right.\nonumber\\
&+&\left.\frac{i}{\hbar}\int^{\,t}\mathrm{d}t^\prime
L(t^\prime)\right],
\end{eqnarray}
where $\chi_N(y)$ is the eigenfunction of the quantum harmonic
oscillator, $\varepsilon_N=\hbar\omega_0\left(N+{1}/{2}\right)$ is
the energy spectrum of the oscillator, $N=0,1,2,...$ is the number
of Landau level, $\omega_0=|e|B/m_e$ is the cyclotron frequency,
$$\zeta(t)=\frac{eE\sin\omega t}{m_e(\omega_0^2-\omega^2)}$$ is the
trajectory of the driven classical oscillator, and
$$L(t)=\frac{m_e\dot{\zeta}^2(t)}{2}-\frac{m_e\omega_0^2\zeta^2(t)}{2}+eE\zeta(t)\sin\omega t$$
is the Lagrangian of the classical oscillator.

It should be noted that the field-induced terms in the wave
functions (\ref{psi})--(\ref{phi}) do not depend on the Landau
level number, $N$. This means that the dressing field does not
change the structure of Landau levels. However, the dressing field
produces exponential phase shifts in the wave functions
(\ref{psi})--(\ref{phi}). In the absence of a magnetic field,
similar phase shifts strongly effect on transport characteristics
of dressed 2DEG via the renormalization of electron scattering
\cite{Kibis_14,Morina_15}. Since the phase shifts in Eqs.
(\ref{psi})--(\ref{phi}) depend on both the dressing field and the
magnetic field, one can expect that magnetotransport properties of
2DEG will be renormalized by the dressing field as well. In order
to describe this renormalization accurately, we have to solve the
scattering problem for the dressed electron states
(\ref{psi})--(\ref{phi}).

Let an electron interact with scatterers in the presence of the
same fields, $\mathbf{A}_0$ and $\mathbf{A}_t$. Then the wave
function of the electron, $\Psi(\mathbf{r},t)$, satisfies the
Schr\"odinger equation
\begin{equation}\label{SE}
i\hbar\frac{\partial\Psi(\mathbf{r},t)}{\partial t}=[\hat{\cal
H}_e+U(\mathbf{r})]\Psi(\mathbf{r},t),
\end{equation}
where $U(\mathbf{r})$ is the total scattering potential of 2DEG
arisen from macroscopically large number of scatterers. Since the
wave functions (\ref{psi}) at any time $t$ coincide with the
eigenfunctions of quantum harmonic oscillator, they form the
complete basis. Therefore, one can seek solutions of the
Schr\"odinger equation (\ref{SE}) as an expansion
\begin{equation}\label{P}
\Psi(\mathbf{r},t)=\sum_{j}a_{j}(t)\psi_{j}(\mathbf{r},t),
\end{equation}
where the different indices $j$ correspond to the different sets
of all quantum numbers ($p_x$ and $N$) describing electron states
of the considered system. It should be stressed that
Eqs.~(\ref{psi})--(\ref{phi}) describe exact wave functions of a
dressed electron. Therefore, the using of the complete basis
(\ref{psi}) in the expansion (\ref{P}) takes into account the
interaction between the electron and the dressing field in full,
i.e. non-perturbatively. As to the electron transition from a
state $j$ to a state $j^{\,\prime}$ due to the potential
$U(\mathbf{r})$, we will describe this scattering process within
the conventional perturbation theory.

Let an electron be in the state $j$ at the time $t=0$ and,
correspondingly, $a_{j^{\,\prime}}(0)=\delta_{j^{\,\prime},j}$.
Substituting the expansion (\ref{P}) into the Schr\"odinger
equation (\ref{SE}) and restricting the accuracy by the first
order of the perturbation theory (the Born approximation), we can
write the amplitude of scattering to the state $j^{\,\prime}$ as
\begin{equation}\label{ak}
a_{j^{\,\prime}}(t)=-\frac{i}{\hbar}\int^{\,t}_0\mathrm{d}t\int_S\mathrm{d}^2\mathbf{r}\,\,\psi_{j^{\,\prime}}^\ast(\mathbf{r},t)U(\mathbf{r})\psi_{j}(\mathbf{r},t),
\end{equation}
where the integration should be performed over the 2DEG area,
$S=L_xL_y$. Applying the Jacobi-Anger expansion,
$$e^{iz\cos\theta}=\sum_{n=-\infty}^{\infty}i^nJ_n(z)e^{in\theta},$$
to transform the time-dependent exponential terms in the wave
functions (\ref{psi})--(\ref{phi}), we arrive from the scattering
amplitude (\ref{ak}) to the scattering probability
\begin{align}\label{wk}
&|a_{j^{\,\prime}}(t)|^2=\frac{\left|U_{j^{\,\prime}j}\right|^2}{\hbar^2}
\Bigg|\sum_{n=-\infty}^{\infty}i^nJ_n\left(\frac{eE[y^\prime_0-y_0]\omega_0^2}{\hbar\omega[\omega_0^2-\omega^2]}\right)\,
\nonumber\\
&\times\,
e^{i(\varepsilon_{j\prime}-\varepsilon_j+n\hbar\omega)t/{2\hbar}}
\int_{-t/2}^{\,t/2}\mathrm{d}t^\prime
\,e^{i(\varepsilon_{j^{\prime}}-\varepsilon_j
+n\hbar\omega)t^\prime/\hbar}\Bigg|^2,
\end{align}
where
\begin{equation}\label{U}
U_{j^{\,\prime}j}=\left\langle\varphi_{j^{\,\prime}}(\mathbf{r})
\left|U(\mathbf{r})\right|\varphi_{j}(\mathbf{r})\right\rangle
\end{equation}
is the matrix element of the scattering between the ``bare''
electron eigenstates,
$$\varphi_j(\mathbf{r})=\frac{e^{ip_xx/\hbar}}{\sqrt{L_x}}\chi_N(y),$$
which satisfy the Schr\"odinger equation with the Hamiltonian
(\ref{H2DEG}) in the absence of the dressing field
$(\mathbf{A}_t=0)$. Since the integral in Eq.~(\ref{wk}) for long
time $t\rightarrow\infty$ turns into the delta function, the
scattering probability (\ref{wk}) can be rewritten as
\begin{eqnarray}\label{wkk}
|a_{\mathbf{k}^\prime}(t)|^2&=&4\pi^2\left|U_{j^{\,\prime}j}\right|^2
\sum_{n=-\infty}^{\infty}J_n^2\left(\frac{eE[y^\prime_0-y_0]\omega_0^2}{\hbar\omega[\omega_0^2-\omega^2]}\right)\nonumber\\
&\times&\delta^2(\varepsilon_{j^{\,\prime}}-\varepsilon_j+n\hbar\omega).
\end{eqnarray}
To transform square delta functions in Eq.~(\ref{wkk}), we can
apply the conventional procedure,
$$\delta^2(\varepsilon)=\delta(\varepsilon)\delta(0)
=\frac{\delta(\varepsilon)}{2\pi\hbar}\lim_{t\rightarrow\infty}
\int_{-t/2}^{\,t/2}e^{i0\times
t^\prime/\hbar}dt^\prime=\frac{\delta(\varepsilon)t}{2\pi\hbar}.$$
Then the probability of the electron scattering between the states
$j$ and $j^{\,\prime}$ per unit time is
\begin{eqnarray}\label{W}
w_{j^{\,\prime}j}&=&\frac{\mathrm{d}|a_{j^{\,\prime}}(t)|^2}{\mathrm{d}
t}=
\left|U_{j^{\,\prime}j}\right|^2\sum_{n=-\infty}^{\infty}J_n^2\left(\frac{eE[y^\prime_0-y_0]\omega_0^2}{\hbar\omega[\omega_0^2-\omega^2]}\right)\nonumber\\
&\times&\frac{2\pi}{\hbar}\,\delta(\varepsilon_{j^{\,\prime}}-\varepsilon_j+n\hbar\omega).
\end{eqnarray}
It should be noted that the derivation of
Eqs.~(\ref{ak})--(\ref{W}) is done within the conventional
time-dependent perturbation theory which is extended to the case
of the oscillating basis (\ref{psi}). Physically, this extension
is similar to the scattering theory developed recently for dressed
electron states in various conductors \cite{Kibis_14,Morina_15}.

To avoid the energy exchange between a high-frequency field and
electrons, the field should be purely dressing (nonabsorbable). In
the considered electron system, there are the two mechanism of
absorption of the field by electrons: (i) the resonant absorption
of the field, which corresponds to electron transitions between
different Landau levels; (ii) the collisional absorption of the
field, which corresponds to electron transitions between different
states within the broadened Landau level. To exclude the first
mechanism, the field frequency, $\omega$, should be far from the
resonant frequencies, $n\omega_0$ ($n=1,2,3...$), corresponding to
interlevel electron transitions. To exclude the second mechanism,
the photon energy, $\hbar\omega$, should be much more than the
scattering-induced broadening of Landau levels,
$\Gamma=\hbar/\tau$ (i.e., $\omega\tau\gg1$). Physically, the
terms with $n\neq0$ in Eq.~(\ref{W}) describe the electron
scattering accompanied by the absorption (emission) of $n$
photons. It follows from the aforesaid that these terms can be
neglected if the dressing field is both off-resonant and
high-frequency. Therefore, the only effect of the dressing field
on 2DEG is the renormalization of the probability of elastic
electron scattering within the same Landau level
($\varepsilon_{j^{\,\prime}}=\varepsilon_j$) which is described by
the term with $n=0$ in Eq.~(\ref{W}):
\begin{equation}\label{W0}
w_{j^{\,\prime}j}=
J_0^2\left(\frac{eE[y^\prime_0-y_0]\omega_0^2}{\hbar\omega[\omega_0^2-\omega^2]}\right)w^{(0)}_{j^{\,\prime}j},
\end{equation}
where
\begin{equation}\label{W00}
w^{(0)}_{j^{\,\prime}j}=\frac{2\pi}{\hbar}\left|U_{j^{\,\prime}j}\right|^2\delta(\varepsilon_{j^{\,\prime}}-\varepsilon_j)
\end{equation}
is the probability of scattering of ``bare'' electron.  As
expected, the probabilities (\ref{W0}) and (\ref{W00}) are
identical in the absence of the dressing field ($E=0$). The formal
difference between the scattering probability of dressed electron
(\ref{W0}) and the scattering probability of ``bare'' electron
(\ref{W00}) consists in the Bessel-function factor depending on
both the dressing field and the stationary magnetic field. Just
this factor is responsible for all effects discussed below.
Particularly, the lifetime of dressed electron at the Landau
level, $\tau$, is renormalized by the Bessel function as
\begin{equation}\label{taud}
\frac{1}{\tau}=\sum_{j^{\,\prime}}w_{j^{\,\prime}j}=
\sum_{j^{\,\prime}}J_0^2\left(\frac{eE[y^\prime_0-y_0]\omega_0^2}{\hbar\omega[\omega_0^2-\omega^2]}\right)
w^{(0)}_{j^{\,\prime}j}.
\end{equation}
In order to calculate the lifetime (\ref{taud}), let us rewrite
the delta function,
$\delta(\varepsilon_{j^{\,\prime}}-\varepsilon_j)$, with using the
well-known representation
\begin{equation}\label{delta}
\delta(\varepsilon)=\frac{1}{\pi}\lim_{\Gamma\rightarrow0}
\frac{\Gamma}{\Gamma^2+\varepsilon^2}.
\end{equation}
In the context of the discussed problem, the parameter
$\Gamma=\hbar/\tau$ has the physical meaning of scattering-induced
broadening of Landau level. For the considered case of elastic
scattering within the same Landau level, we can write the delta
function (\ref{delta}) as
$\delta(\varepsilon_{j^{\,\prime}}-\varepsilon_j)\approx1/(\pi\Gamma)$
and, therefore, Eq.~(\ref{taud}) takes the form
\begin{equation}\label{taudd}
\frac{1}{\tau}=
\left[\frac{2}{\hbar^2}\sum_{j^{\,\prime}}J_0^2\left(\frac{eE[y^\prime_0-y_0]\omega_0^2}{\hbar\omega[\omega_0^2-\omega^2]}\right)
\left|U_{j^{\,\prime}j}\right|^2\right]^{1/2},
\end{equation}
where the summation is performed over electron states
$j^{\,\prime}$ within the same Landau level. To calculate the
lifetime (\ref{taudd}), let us approximate the scattering
potential using the model of delta-function scatterers,
$$U\left(\mathbf{r}\right)=\sum_{i=1}^{N_s} U_0
\delta\left(\mathbf{r}-\mathbf{r}_i\right),$$ which is commonly
used to describe electronic transport in various two-dimensional
systems \cite{Ando_74,Ando_82,Burkov_2004,Burkov_2010}. Assuming
that the scatterers to be distributed randomly and the total
number of scatterers, $N_s$, to be macroscopically large, we can
obtain from Eq.~(\ref{taudd}) the final expression for the
electron lifetime at the $N$-th Landau level,
\begin{align}\label{tauf}
&\frac{1}{\tau}=\sqrt{\frac{
n_sU_0^2}{\pi l_0^2\hbar^2}}\,\times\nonumber\\
&\left[\iint_{-\infty}^{\infty}\chi_N^2(y^\prime)\chi_N^2(y+y^\prime)
J_0^2\left(\frac{eEy\omega_0^2}{\hbar\omega[\omega_0^2-\omega^2]}\right)
\mathrm{d}y\,\mathrm{d}y^\prime\right]^{1/2},
\end{align}
where $n_s=N_s/S$ is the density of scatterers per unit area of
2DEG, and $l_0=\sqrt{\hbar/|e|B}$ is the magnetic length. The
argument of the Bessel function in the integrand of
Eq.~(\ref{tauf}) is the dimensionless parameter which describes
the ratio of the characteristic energy of the electron-field
interaction and the photon energy. Physically, it describes the
strength of electron-photon coupling in the considered
electron-field system. Since the dressing field, $E$, leads to
decreasing the Bessel function, the scattering time, $\tau$,
increases due to the field. Magnetoelectronic effects following
from this increasing are discussed below.

\section{Magnetoelectronic characteristics of dressed 2DEG}
Since the scattering time (\ref{tauf}) depends on the dressing
field, the scattering-induced broadening of Landau levels,
$\Gamma=\hbar/\tau$, is also affected by the field. In order to
describe the broadening accurately, it is convenient to rewrite
Eq.~(\ref{tauf}) in the dimensionless form,
\begin{align}\label{gammat}
&\frac{\Gamma^{(N)}}{\Gamma_0}=\nonumber\\
&\left[\iint_{-\infty}^{\infty}\chi_N^2(y^\prime)\chi_N^2(y+y^\prime)
J_0^2\left(\frac{eEy\omega_0^2}{\hbar\omega[\omega_0^2-\omega^2]}\right)
\mathrm{d}y\,\mathrm{d}y^\prime\right]^{1/2},
\end{align}
where $\Gamma^{(N)}=\hbar/\tau$ is the broadening for the Landau
level with the number $N=0,1,2...$, and $\Gamma_0$ is the
broadening of Landau levels in the absence of the dressing field
(natural broadening). It should be noted that Eq.~(\ref{gammat})
does not depend on the density of scatterers, $n_s$, and the
strength of scatterers, $U_0$. Therefore, Eq.~(\ref{gammat})
describes the dependence of the broadening of Landau levels on the
dressing field in the most general form, where the broadening of
``bare'' Landau levels, $\Gamma_0$, should be treated as a
phenomenological parameter which can be found from experiments. In
the absence of the dressing field ($E=0$), the broadening
(\ref{gammat}) is the same for all Landau levels,
$\Gamma=\Gamma_0\propto \sqrt{B}$, in complete agreement with the
conventional theory of magnetoelectronic properties of 2DEG
\cite{Ando_74,Ando_82}. On the contrary, the dressing field leads
to the different broadening (\ref{gammat}) for different Landau
levels (see Fig.~2).
\begin{figure}[th]
\includegraphics[width=0.48\textwidth]{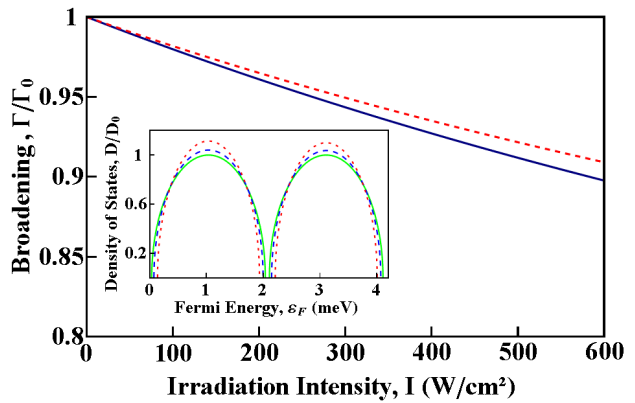}
\caption{(Color online) The dependence of the broadening of Landau
levels, $\Gamma$, on the irradiation intensity, $I$, for the
lowest two Landau levels with the numbers $N=0$ (solid line) and
$N=1$ (dashed line) in a GaAs-based quantum well at the magnetic
field $B=1.2$ T, the irradiation frequency $\omega=2\cdot10^{12}$
rad/s, and the natural broadening $\Gamma_0=1$ meV. The insert
shows the density of electron states, $D$, in the absence of the
irradiation (solid line) and for the irradiation intensities
$I=200$~W/cm$^2$ (dashed line), $I=600$~W/cm$^2$ (dotted
line).}\label{Fig3}
\end{figure}
As to the density of electron states, it is described by the
expression \cite{Ando_74,Ando_82}
\begin{equation}\label{D}
D(\varepsilon)=D_0\sum_N\frac{\Gamma_0}{\Gamma^{(N)}}
\left[1-\left(\frac{\varepsilon-\varepsilon_N}{\Gamma^{(N)}}\right)^2\right]^{1/2},
\end{equation}
where $D_0=1/(\pi^2l_0^2\Gamma_0)$. Substituting the broadening
(\ref{gammat}) into Eq.~(\ref{D}), one can calculate the density
of states in dressed 2DEG (see the insert in Fig.~2). Since the
dressing field decreases the broadening of Landau levels
(\ref{gammat}), this results in increasing the density of states
at Landau level energies, $\varepsilon=\varepsilon_N$. As a
consequence, all phenomena sensitive to the density of electronic
states (magnetotransport, magneto-optical effects, etc) are
affected by the dressing field. Particularly, the longitudinal
magnetoconductivity of 2DEG at the temperature $T=0$ is described
by the conventional expression \cite{Ando_74,Ando_82},
\begin{equation}\label{cond}
\sigma_{xx}\approx\sigma_0\left(N+\frac{1}{2}\right)\left[1-\left(\frac{\varepsilon-\varepsilon_N}{\Gamma^{(N)}}\right)^2\right],
\end{equation}
where $\sigma_0=e^2/\pi^2\hbar$ is the elementary conductivity,
and $N$ is the number of Landau level at the Fermi energy.
Substituting the broadening (\ref{gammat}) into Eq.~(\ref{cond}),
one can calculate the dependence of the conductivity on the
dressing field. Experimentally, one can change the Fermi energy of
2DEG, $\varepsilon_F$, with a gate voltage. Then we arrive from
Eq.~(\ref{cond}) at the oscillating behavior of the conductivity
(the Shubnikov-de Haas oscillations) plotted in Fig.~3.
\begin{figure}[th]
\includegraphics[width=0.48\textwidth]{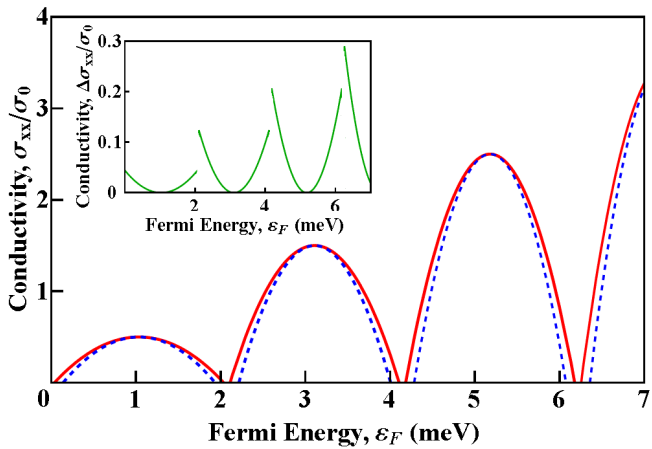}
\caption{(Color online) The dependence of the longitudinal
conductivity, $\sigma_{xx}$, on the Fermi energy, $\varepsilon_F$,
in a GaAs-based quantum well at the magnetic field $B=1.2$ T,
irradiation frequency $\omega=2\cdot10^{12}$ rad/s, and the
natural broadening $\Gamma_0=1$ meV. The solid line describes the
conductivity of unirradiated 2DEG, whereas the dotted one
corresponds to the conductivity at the irradiation intensity
$I=600$~W/cm$^2$. The insert shows the difference of these two
conductivities, $\Delta\sigma_{xx}$.} \label{Fig4}
\end{figure}

It should be stressed that there is the crucial difference between
the considered high-frequency dressing field and the low-frequency
case. Since 2DEG absorbs a low-frequency field, the
multiphoton-assisted scattering of electrons can increase the
longitudinal conductivity \cite{Lei_2003}. Particularly, this
effect was proposed to explain the phenomenon of ``zero resistance
states'' in 2DEG subjected to both a magnetic field and a
low-frequency (microwave) irradiation
\cite{Zudov_2001,Dmitriev_2012}. On the contrary, the considered
high-frequency field cannot be absorbed by 2DEG. The only effect
of the field is the suppression of electron scattering, which
results in decreasing both the broadening of Landau levels and the
longitudinal conductivity (see Figs.~2 and 3). Thus, a
high-frequency irradiation and a low-frequency one lead to
different behavior of the magnetoelectronic properties of 2DEG.

It should be noted  also that the magnetoelectronic effects
induced by a dressing field strongly depends on the kind of
electron dispersion. In Dirac materials with linear electron
dispersion, a dressing field changes the energy distance between
Landau levels and, therefore, modifies all phenomena depending on
the cyclotron frequency \cite{Kibis_16}. On the contrary, in the
considered case of 2DEG with the parabolic electron dispersion, a
dressing field does not change the cyclotron frequency but
influences on the electron scattering within Landau levels.

As to experimental observability of the discussed phenomena, all
dressing effects increase with increasing the intensity of the
dressing field. Particularly, the strong dressing field can turn
the Bessel function in Eq.~(\ref{tauf}) into zero, what
corresponds physically to the field-induced suppression of
electron scattering \cite{Kibis_14}. However, an intense
irradiation can fluidize a semiconductor quantum well. To avoid
the fluidizing, it is reasonable to use narrow pulses of a strong
dressing field. This well-known methodology has been elaborated
long ago and commonly used to observe various dressing effects ---
particularly, modifications of energy spectrum of dressed
electrons arisen from the optical Stark effect
--- in semiconductor structures (see, e.g., Refs.~\onlinecite{Joffre_1988_1,Joffre_1988_2,Lee_1991}). Within this approach,
giant dressing fields (up to GW/cm$^2$) can be applied to
semiconductor structures.

\section{Conclusions}
Summarizing the aforesaid, we can conclude that a strong
high-frequency electromagnetic field (dressing field) decreases
the electron scattering between different cyclotron orbits within
the same Landau level. As a consequence, the field decreases the
scattering-induced broadening of Landau levels in 2DEG. This
results in the field-induced modification of various
magnetoelectronic properties depending on the density of electron
states (particularly, magnetotransport characteristics of 2DEG).
Therefore, a dressing field can be considered as a perspective
tool to manipulate the magnetoelectronic properties of various
two-dimensional nanostructures. Since such nanostructures serve as
a basis for nanoelectronic devices, the developed theory opens a
way for optical control of their magnetoelectronic
characteristics.

\begin{acknowledgements}
The work was partially supported by FP7 IRSES project QOCaN, FP7
ITN project NOTEDEV, Rannis project BOFEHYSS, Singaporean Ministry
of Education under AcRF Tier 2 grant MOE2015-T2-1-055, RFBR
project 14-02-00033 and Russian Ministry of Education and Science.
I.A.S. thanks the support of 5-100 program of Russian Federal
Government.
\end{acknowledgements}

\end{document}